# Cavitation of water by volume-controlled stretching

Peng Wang[1], Wei Gao[2], Justin Wilkerson[2], Kenneth M. Liechti[1] and Rui Huang[1*]

[1]*Department of Aerospace Engineering and Engineering Mechanics, University of Texas at Austin, Austin, TX 78712*

[2]*Department of Mechanical Engineering, University of Texas at San Antonio, San Antonio, TX 78249*

## Abstract

A liquid subjected to negative pressure is thermodynamically metastable. Confined within a small volume, negative pressure can build up until cavities form spontaneously. The critical negative pressure for cavitation in water has been theoretically predicted to be in the range of -100 to -200 MPa at room temperature, whereas values around -30 MPa have been obtained by many experiments. The discrepancy has yet to be resolved. In this study we perform molecular dynamics simulations to study cavitation of water under volume controlled stretching. It is found that liquid water exhibits a nonlinear elastic compressibility (or stretchability) under hydrostatic tension and remains stable within the confined volume until spontaneous cavitation occurs at a critical strain. Subsequently, as the volume-controlled stretching continues, the cavity grows stably and the hydrostatic tension decreases continuously until the box volume is large enough for another transition to form a water droplet. A modified nucleation theory is proposed to predict the critical condition for cavitation. In particular, a strong dependence of the critical strain and stress for cavitation on the initial liquid volume is predicted by the modified nucleation theory, which may offer a possible explanation for the discrepancies in the values of the critical negative pressure obtained from experiments.

Keywords: water, cavitation, phase transition, negative pressure

---

[*] Corresponding author. Email: ruihuang@mail.utexas.edu.

# 1. Introduction

Liquid water can be stretched under negative pressure (or hydrostatic tension) in a metastable state until cavitation [1]. Classical nucleation theory predicts a critical negative pressure for cavitation to be around -140 MPa [2] or -160 MPa [3] around room temperature (~300 K). Experimentally, a critical negative pressure of -140 MPa was reached at 315 K using the microscopic Berthelot tube method [4-6], by cooling a micrometer sized water droplet confined in a quartz crystal so that the liquid, which sticks to the hydrophilic wall, follows an isochore (constant density path) to the point of cavitation. Many other experimental methods have been used to study cavitation of water, such as acoustic cavitation [7, 8], shock wave techniques [9], and the method of artificial trees [10]. However, the critical negative pressure values from these experiments are typically around -30 MPa [11], far from the theoretical prediction. The reason for the apparent discrepancy remains unknown. While extensive studies have been devoted to supercooled water (metastable with respect to crystallization or vitrification) [11-14], a unified description of water in its stable and metastable states remains elusive, and the knowledge about liquid water under negative pressure (metastable with respect to vaporization) is still in its infancy. It is thus believed that the extreme mechanics of water under negative pressure including the study of cavitation can provide essential insights for establishing the metastable phase diagram of water and explaining several thermodynamic anomalies of water [11].

In nature, negative pressure in water occurs routinely in the xylem of trees [15], where cavitation breaks the continuity of water columns and hence the water supply to transpiring leaves [16]. A similar situation occurs with water filling a void in an elastomer [17], which is in a state of hydrostatic tension due to osmosis. Moreover, liquid water in a narrow capillary could be under negative pressure as a result of the Laplace pressure with a curved liquid–vapor interface [18], such as water in soil. A related phenomenon is the capillary bridge at the interface between two solids, a common feature of wet adhesion [19-21]. As shown in a recent study [21], separation of a graphene membrane from a wet surface could stretch the water film in between to the point of cavitation with a critical negative pressure of about -90 MPa, depending on the graphene-water interactions.

In this study, we conduct molecular dynamics (MD) simulations to study cavitation of water under volume-controlled conditions (Section 2). Motivated by the MD simulations, we propose a modified nucleation theory by taking into account the nonlinear elastic compressibility of liquid water (Section 3). The numerical and theoretical results are compared and discussed in Section 4 to elucidate the effects of relaxation time, initial volume, and temperature on cavitation as well as a complete pathway of phase transition by volume-controlled stretching, from the stable liquid phase to the stable vapor phase of water through a series of metastable states (including cavitation and droplet).

## 2. MD Simulation

We perform classical MD simulations using LAMMPS [22]. Many different empirical potentials have been proposed for water, such as ST2 [23], SPC/E [24], TIP4P [25], TIP5P [26], TIP4P/2005 [27], and polarizable AMOEBA [28, 29] models. In this study we use the TIP4P/2005 model, which accurately describes the surface tension of water over the whole range of temperatures from the triple point to the critical temperature [30]. In the TIP4P/2005 model, each water molecule has four interaction sites, including a massless M-site located coplanar with the oxygen (O) and hydrogen (H) sites on the bisector of the H-O-H bond angle. The O-H bond length and the H-O-H bond angle are fixed as 0.9572 Å and 104.52°, respectively. The intermolecular pair potential has two contributions, a Lennard-Jones (LJ) term and an electrostatic part. The oxygen site carries no charge, but contributes to the LJ term:

$$U_{LJ} = 4E_{OO}\left[\left(\frac{L_{OO}}{r_{OO}}\right)^{12} - \left(\frac{L_{OO}}{r_{OO}}\right)^{6}\right], \quad (1)$$

where $r_{OO}$ is the distance between the oxygen sites of two water molecules, $L_{OO} = 3.1589\,\text{Å}$ and $E_{OO} = 0.00803\,\text{eV}$. The H and M sites are charged ($q_H = 0.5564e$ and $q_M = -2q_H = -1.1128e$), but do not contribute to the LJ term. The electrostatic part of the interaction potential between two water molecules is

$$U_e = \sum_{i,j(i\neq j)} \frac{k_e q_i q_j}{r_{ij}}, \quad (2)$$

where the summation is taken over all pairs of charged sites, $r_{ij}$ is the distance between two charged sites, and $k_e$ is the electrostatic constant. The total potential energy of the system is the sum of the pair potentials between all molecules. The cutoff distance is set to be 13 Å for both the LJ and electrostatic interactions. The electrostatic interactions are computed by using the particle-particle particle-mesh (PPPM) algorithm [31] as implemented in LAMMPS. The time integration scheme closely follows the time-reversible measure-preserving Verlet and rRESPA integrators derived by Tuckerman et al. [32], with a time step of 1 fs.

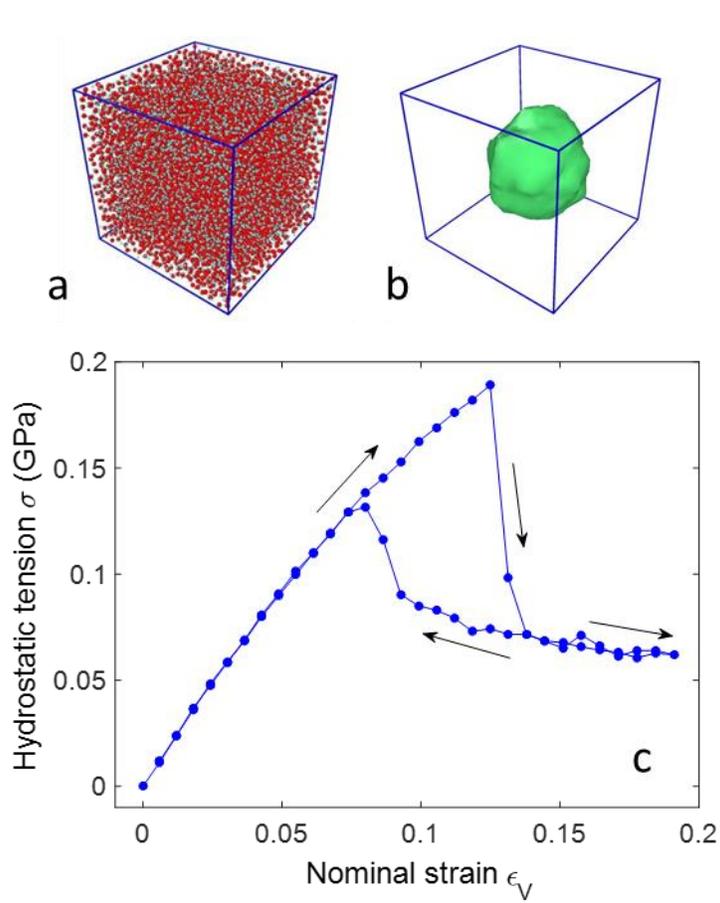

**Figure 1**. MD simulation of water cavitation under volume-controlled stretching. (a) A simulation box of 8000 water molecules ($L_0 = 6.22$ nm); (b) Cavity surface at $\varepsilon_V = 0.191$, constructed by the alpha-shape method [33] with a virtual probe sphere of radius 0.4 nm using OVITO [34]. (c) Calculated hydrostatic tension ($\sigma = -p$) versus the nominal volumetric strain ($\varepsilon_V$), subjected to a loading-unloading cycle at $T = 300$ K.

We consider in our simulations a cubic box of water molecules with periodic boundary conditions, as shown in Fig. 1. The temperature is controlled by using a Nose-Hoover thermostat. Each simulation starts with a constant pressure isotropic relaxation (NPT) under the atmosphere pressure ($p = 1$ bar or 0.1013 MPa) to equilibrate the system for 1 ns. The equilibrium box size is obtained as:

$$L_0 = \frac{1}{3}\langle L_x + L_y + L_z \rangle_t, \tag{3}$$

where $L_x, L_y, L_z$ are the linear dimensions of the simulation box and $\langle \cdot \rangle_t$ denotes the time average of the enclosed quantity over the period of relaxation. Next, the simulation is switched to NVT ensemble to simulate volume-controlled stretching while the pressure is calculated and may become negative. A incremental linear strain of $\Delta \varepsilon_L = 2 \times 10^{-3}$ is applied by stretching the simulation box in three dimensions simultaneously with $\Delta L = L_0 \Delta \varepsilon_L$, followed by NVT relaxation for 0.2 ns. The corresponding pressure is calculated at each strain level by averaging the diagonal components of the virial stress tensor, namely

$$p = -\frac{1}{3} \sum_{k=1}^{3} \sigma_{kk} . \tag{4}$$

The virial stress [35] is calculated as

$$\boldsymbol{\sigma} = \left\langle -\frac{1}{2L^3} \sum_{\substack{i,j \\ i \neq j}} \mathbf{F}_{ij} \otimes (\mathbf{r}_i - \mathbf{r}_j) - \frac{1}{L^3} \sum_i m_i \mathbf{v}_i \otimes \mathbf{v}_i \right\rangle_t, \tag{5}$$

where $\mathbf{F}_{ij}$ is the force vector between two interaction sites (O, H, or M), $\mathbf{r}_i$ is the position vector, $\mathbf{v}_i$ is the velocity vector, $m_i$ is the atomic mass associated with O and H sites only. Note that Eq. (5) gives the average stress over the volume ($V = L^3$) including the volume of cavity (if exists).

As shown in Fig. 1 for $L_0 = 6.22$ nm at $T = 300$ K, the hydrostatic tension is nearly zero after NPT relaxation ($\sigma \sim -0.1$ MPa), and the molecular number density is 33.2 nm$^{-3}$, close to the theoretical value of 33.4 nm$^{-3}$ for liquid water. As the volumetric strain ($\varepsilon_V = L^3/L_0^3 - 1$) increases, the hydrostatic tension first increases and then drops abruptly. The initial bulk modulus of water is obtained from the slope ($d\sigma/d\varepsilon_V$) of the hydrostatic stress-strain diagram at small strains, which is 2.09 GPa, reasonably close to the value (~2.2 GPa) measured in experiments [36]. As the strain increases, the tangent bulk modulus decreases, indicating a nonlinear elastic behavior of liquid water under tension. The hydrostatic tension reaches a maximum of 189 MPa at $\varepsilon_V = 0.124$, after which the stress drops abruptly. It is found that the sudden stress drop is associated with onset of cavitation in the water box, which leads to a nearly spherical cavity as shown in Fig. 1 (top right). Hence, the MD simulation predicts a critical negative pressure for cavitation at -189 MPa under volume-controlled stretching at 300 K. Subsequently, as the strain increases further, the cavity grows larger in size and the hydrostatic tension decreases slowly. Upon unloading by decreasing the box volume, the cavity shrinks and the stress increases. Eventually, the cavity disappears at a lower critical stress (~140 MPa), and further unloading follows the loading curve back to the initial state at zero strain. A hysteresis loop is thus observed in the hydrostatic stress-strain diagram after the loading/unloading cycle, indicating energy dissipation associated with the cavitation process.

## 3. Modified Nucleation Theory

Classical nucleation theory has been commonly adopted to understand cavitation of water under negative pressure [1-3, 13]. At a given pressure and temperature, the net work associated with formation of a spherical cavity of radius $R$ in the liquid phase is written as

$$W(R;p,T) = \frac{4}{3}\pi R^3 [p - p_0(T)] + 4\pi R^2 \gamma(T), \tag{6}$$

where $p_0(T)$ is the equilibrium vapor pressure in the cavity and $\gamma(T)$ is the surface tension of the liquid/vapor interface; both are functions of temperature. For $p < p_0(T)$, the first term on the right hand side of Eq. (6) is negative, whereas the second term is always positive. The competition of the two terms defines a critical cavity radius:

$$R_c(p,T) = \frac{2\gamma(T)}{p_0(T) - p}. \tag{7}$$

As a result, when $p < p_0(T)$, smaller cavities ($R < R_c(p,T)$) disappear and larger cavities ($R > R_c(p,T)$) grow. The corresponding energy barrier is:

$$E_b(p,T) = \frac{16\pi}{3} \frac{\gamma(T)^3}{[p_0(T) - p]^2}. \tag{8}$$

Then, by the kinetic theory of nucleation, the rate of cavitation is proportional to $\exp[-E_b/(k_B T)]$, where $k_B$ is Boltzmann constant. As such, the critical negative pressure for cavitation was predicted as the pressure to form the first cavity within a specific time $\tau$ [2]:

$$p_c(T) = p_0(T) - \left(\frac{16\pi}{3} \frac{\gamma(T)^3}{k_B T \ln(\tau N k_B T/h)}\right)^{1/2}, \tag{9}$$

where $N$ is the number of molecules and $h$ is Planck's constant. Interestingly, the critical pressure by Eq. (9) was found to be extremely insensitive to the particular time $\tau$ [2], and it was suggested to take the critical pressure as that corresponding to formation of one cavity per second. Taking $\tau = 1\,\text{s}$ and $N = 6.022 \times 10^{23}$ for 1 mole of water molecules along with $\gamma = 0.072$ N/m for liquid water at $T = 300$ K, the critical negative pressure for cavitation was predicted to be -134 MPa [2].

A slightly different version of the classical nucleation theory [1, 18, 37] considered the probability to form a cavity by thermal fluctuation and defined the cavitation pressure as the pressure when the probability reaches 50%. This led to a similar prediction as Eq. (9):

$$p_c(T) = p_0(T) - \left(\frac{16\pi}{3} \frac{\gamma(T)^3}{k_B T \ln(\Gamma_0 V \tau / \ln 2)}\right)^{1/2}, \tag{10}$$

where $V$ is the volume of liquid and $\Gamma_0$ is a kinetic pre-factor for the nucleation rate. The pre-factor, which was not given precisely, was estimated as the product of the thermal frequency ($k_B T / h$) and the density of independent nucleation sites ($\sim 1/R_c^3$) [37]. With $V = 1000\,\mu\text{m}^3$ and $\tau = 1\,\text{s}$, Eq. (10) predicts a cavitation pressure at -168 MPa at 293 K [1]. It can be shown that the prediction by Eq. (10) is insensitive to the volume, ranging from around -220 MPa for $V = 1\,\text{nm}^3$ to -130 MPa for $V = 1\,\text{m}^3$.

Motivated by the MD simulations as described in Section 2, we propose a modified nucleation theory. In particular, we take into account the finite compressibility of liquid water and consider cavitation under the volume-controlled condition (instead of pressure control). In this way, the cavitation process is stabilized, which allows the prediction of subsequent cavity growth in addition to the critical negative pressure. Let $V_0$ be the initial volume of the liquid water in its stable phase at a given pressure (e.g, $p = 1\,\text{bar}$) and temperature (e.g., $T = 300$ K). Stretch the volume to $V > V_0$ under isothermal condition. Assuming a spherical cavity of radius $R$ in $V$ ($R = 0$ if no cavity), we write the free energy of the system as

$$\Phi = V_0 U_L(\varepsilon) + 4\pi R^2 \gamma - \frac{4}{3}\pi R^3 p_0, \tag{11}$$

where $U_L(\varepsilon)$ is the elastic strain energy density of the liquid as a function of the volumetric strain $\varepsilon$. To obtain a functional form of $U_L(\varepsilon)$, we observe from the MD simulation (Fig. 1) that the hydrostatic stress before cavitation can be written as a function of the volumetric strain:

$$\sigma = K_1 \varepsilon + K_2 \varepsilon^2, \tag{12}$$

where $K_1$ is the linear bulk modulus and $K_2$ is the second-order modulus for the nonlinear behavior at relatively large strains. Both moduli are temperature dependent as discussed later. For $T = 300$ K, we fit the hydrostatic stress-strain diagram from the MD simulation to obtain $K_1 = 2.09$ GPa and $K_2 = -4.62$ GPa, as shown in Fig. 2a (Branch I). By integrating Eq. (12) with respect to the strain, we obtain the strain energy density function as

$$U_L(\varepsilon) = \frac{1}{2} K_1 \varepsilon^2 + \frac{1}{3} K_2 \varepsilon^3. \tag{13}$$

Let $\varepsilon_V = V/V_0 - 1$ be the nominal strain. With a spherical cavity of radius $R$ in $V$, the volume of liquid water is $V - \frac{4}{3}\pi R^3$ and hence the volumetric strain in the liquid is

$$\varepsilon = \varepsilon_V - \frac{4\pi R^3}{3V_0}. \tag{14}$$

Substituting Eq. (14) into Eq. (13) and then into Eq. (11), we obtain

$$\Phi = V_0 U_L(\varepsilon_V) + \Delta\Phi(R, \varepsilon_V), \tag{15}$$

where

$$\Delta\Phi(R,\varepsilon_V) = 4\pi R^2 \gamma \left[ 1 - \frac{1}{3}\left(\varepsilon_V + \frac{K_2}{K_1}\varepsilon_V^2\right)\frac{R}{l} + \frac{2\pi l^3}{9V_0}\left(1 + \frac{2K_2}{K_1}\varepsilon_V\right)\left(\frac{R}{l}\right)^4 - \frac{16\pi^2 l^6 K_2}{81 V_0^2 K_1}\left(\frac{R}{l}\right)^7 \right] - \frac{4}{3}\pi R^3 p_0 \tag{16}$$

with a length scale $l = \gamma/K_1$. The first term on the right-hand side of Eq. (15) gives the free energy without cavitation, while the second term is the change of free energy with cavitation in the same total volume.

Consider a cubic box of liquid water with $L_0 = 6.22$ nm and $V_0 = L_0^3$ at $T = 300$ K (same as the MD simulation in Fig. 1). Stretched to a nominal strain $\varepsilon_V$, the change of free energy in Eq. (16), normalized by $4\pi l^2 \gamma$, is plotted as a function of the dimensionless cavity size $R/l$ in Fig. 2b. Here, we take $\gamma = 0.0693$ N/m as the surface tension of water at 300 K as predicted by the TIP4P/2005 model [30], along with the bulk moduli, $K_1 = 2.09$ GPa and $K_2 = -4.62$ GPa; the vapor pressure ($p_0$) is negligible and thus ignored hereafter. When the nominal strain is small ($\varepsilon_V < 0.094$), the free energy function has a single minimum at $R = 0$, and thus the liquid water is uniformly stretched without any cavity and remains stable under the NVT condition although the pressure may become negative. When the nominal strain exceeds a threshold value ($\sim 0.094$), a local minimum of the free energy function appears at a finite cavity radius ($R > 0$). The free energy at the local minimum decreases with increasing nominal strain. When the nominal strain exceeds a critical value ($\sim 0.111$), the free energy at the second minimum ($R > 0$) becomes lower than the free energy at the first minimum ($R = 0$). As a result, the uniformly stretched liquid water becomes metastable with respect to formation of a cavity, whereas the state with a finite cavity ($R > 0$) is stable under the NVT condition. Therefore, by volume-controlled stretching, cavitation is predicted as a first-order phase transition from the homogeneous liquid phase (stretched but no cavity) to the cavitated state with coexisting liquid and vapor phases within the volume.

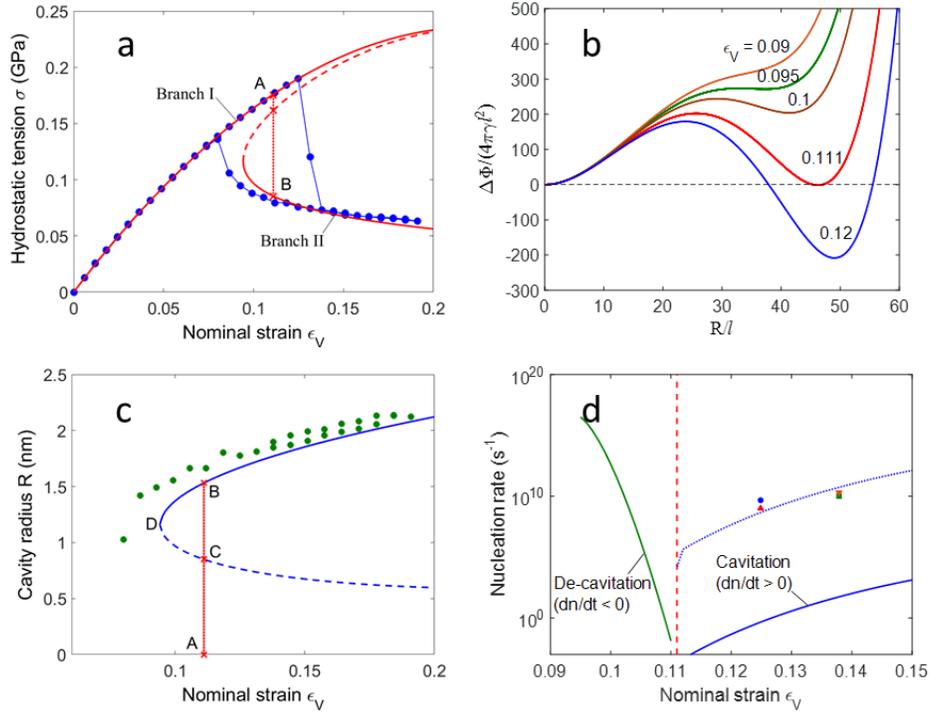

**Figure 2**. Prediction of cavitation by a modified nucleation theory: (a) Hydrostatic stress-strain diagram, in comparison with MD simulation (filled symbols). (b) Free energy function under different nominal strains. (c) Cavity radius as a function of the nominal strain, in comparison with MD simulation (filled symbols). (d) Nucleation rate for cavitation and de-cavitation. The dashed lines in (a) and (c) are the unstable branches. The vertical lines in (a) and (c) indicate the transition

(from A to B) at the critical strain. The dotted line in (d) is obtained by shifting the solid line up by a factor of $10^9$, to compare with $1/\tau_R$ in MD simulations (symbols).

The critical strain for cavitation can be predicted by setting the free energy values at the two minima to be equal. First, the radius of cavity is obtained as a function of the nominal strain (Fig. 2c) by setting $\partial \Phi / \partial R = 0$ and $\partial^2 \Phi / \partial R^2 > 0$ for the energy minimum. It can be shown that the cavity radius thus obtained satisfies the Young-Laplace equation as expected, namely

$$R(\varepsilon_V) = \frac{2\gamma}{\sigma}, \qquad (17)$$

where $\sigma$ is the hydrostatic tension in liquid (after cavitation) and can be obtained from Eq. (12) with the strain in Eq. (14). Next, by inserting $R(\varepsilon_V)$ into Eq. (16) and setting $\Delta \Phi = 0$, we obtain an equation that can be solved to predict the critical strain for cavitation. As shown in Fig. 2c, a discontinuous transition (from point A to point B) is predicted at the critical strain ($\varepsilon_c \sim 0.111$), whereas the radius for the metastable cavity is obtained before the critical strain ($0.094 < \varepsilon < 0.111$). Also shown in Fig. 2c is an unstable branch (dashed line) corresponding to the maximum free energy between the two energy minima in Fig. 2b, which may be considered as the critical cavity radius ($R = R_c$) under the volume-controlled stretching.

At the critical strain ($\varepsilon_c \sim 0.111$), the hydrostatic tension changes abruptly before and after cavitation (from A to B in Fig. 2a). Before cavitation, the hydrostatic stress can be obtained directly from Eq. (12) with $\varepsilon = \varepsilon_c$. This gives a critical stress of 175 MPa, slightly lower than that from the MD simulation. After cavitation, the volumetric strain in liquid is less than the nominal strain as given in Eq. (14), with which the hydrostatic stress can be calculated. Alternatively, the stress after cavitation can also be calculated by Eq. (17). To compare with the virial stress in Eq. (5) by the MD simulation, the volume-averaged hydrostatic stress is calculated as a function of the nominal strain as

$$\bar{\sigma}(\varepsilon_V) = \sigma(\varepsilon)\left[1 - \frac{4\pi R^3}{3V_0(1+\varepsilon_V)}\right]. \qquad (18)$$

As shown in Fig. 2a, the hydrostatic stress-strain diagram consists of two branches. Before cavitation (branch I), the stress is well fit by Eq. (12). After cavitation (branch II), the stress calculated by Eq. (18) compares closely with the MD results. In addition, an unstable branch (dashed line) is shown in between, corresponding to the critical cavity radius (dashed line in Fig. 2c). The unstable branch connects with branch II at a threshold strain ($\sim 0.094$) with infinite slope and approaches branch I asymptotically at increasingly large strain. With Eq. (17) and Eq. (18), the cavity radius in the MD simulation can be estimated by the calculated virial stress. As shown in Fig. 2c, the estimated radius agrees with the prediction reasonably well.

Furthermore, we note that in the MD simulation cavitation occurs at a strain greater than the predicted critical strain and disappears at a smaller strain during unloading. This may be qualitatively understood by a kinetic theory of nucleation [13]. Under the NVT condition, the free energy function has two local minima when the nominal strain is greater than a threshold value ($\sim 0.094$) as shown in Fig. 2b. In this case, the net rate of cavity formation can be written as:

$$\frac{dn}{dt} = \Gamma\left[\exp\left(-\frac{\Delta\Phi_{max}}{k_B T}\right) - \exp\left(-\frac{\Delta\Phi_{max} - \Delta\Phi_{min}}{k_B T}\right)\right], \quad (19)$$

where $\Delta\Phi_{max}$ is the energy difference between the homogeneous state ($R = 0$) and the state with the locally maximum free energy ($R = R_c$) and $\Delta\Phi_{min}$ is the energy difference between the two energy minima. Thus, $\Delta\Phi_{max}$ is the energy barrier to form a cavity and $\Delta\Phi_{max} - \Delta\Phi_{min}$ is the energy barrier for the cavity to disappear, both varying with the nominal strain as shown in Fig. 2b. The kinetic pre-factor $\Gamma$ is not known precisely. Nevertheless, we plot the nucleation rate as a function of the nominal strain in Fig. 2d by taking $\Gamma = Nk_B T/h$, where $N$ is the number of water molecules in the MD simulation ($N = 8000$). At the critical strain ($\varepsilon_c \sim 0.111$), the net nucleation rate is zero. When $\varepsilon < \varepsilon_c$, $\Delta\Phi_{min} > 0$ and the nucleation rate is negative. On the other hand, when $\varepsilon > \varepsilon_c$, $\Delta\Phi_{min} < 0$ and the nucleation rate is positive. For cavitation to occur within a finite time, a positive nucleation rate is required and hence the cavitation strain observed in the MD simulation is larger than the critical strain. Similarly, during unloading, the nucleation rate must be negative for the cavity to disappear and hence the de-cavitation strain is lower than the critical strain. Quantitatively, however, the nucleation rate by Eq. (19) is too low for a cavity to form within the relaxation time (0.2 ns) at each strain level, possibly due to the rough estimate of the pre-factor $\Gamma$. For unloading, the MD simulation shows that the cavity remains at strain levels below the threshold ($\sim 0.094$), which may suggest that the relaxation time was too short for the cavity to disappear.

However, the negative nucleation rate by Eq. (19) for $\varepsilon < \varepsilon_c$ is very high in magnitude and thus the relaxation time should be sufficient. Therefore, the kinetic theory of nucleation, while qualitatively consistent with the MD simulation, is not quantitatively accurate in predicting the cavitation/de-cavitation strains.

## 4. Results and Discussion

### 4.1 Effect of relaxation time

MD simulations are often limited by computational cost to the cases with unrealistically high loading rates or short relaxation time. With a linear strain increment $\Delta\varepsilon_L = 2\times 10^{-3}$ and a relaxation time $\tau_R = 0.2$ ns at each strain level, the linear strain rate is $10^7 \, \text{s}^{-1}$. By increasing the relaxation time to 1 ns in MD simulations, thus reducing the loading rate by a factor of 5, the hydrostatic stress-strain diagram shows little difference (Fig. 3). On the other hand, if decreasing the relaxation time to 0.1 and 0.05 ns, the onset of cavitation is slightly delayed till a higher strain and correspondingly a higher magnitude for the cavitation pressure. The stress-strain behavior before cavitation and during unloading remains unaffected by changing the relaxation time. Theoretically, the relaxation time may be related to the nucleation rate that is required for a cavity to form, roughly, $dn/dt \sim 1/\tau_R$. Thus, a longer relaxation time requires a lower nucleation rate and correspondingly a smaller critical strain as shown in Fig. 2d, which is qualitatively consistent with the MD simulations in Fig. 3. However, without a quantitatively accurate prediction of the nucleation rate, there is a large gap between the estimated nucleation rate by Eq. (19) with $\Gamma = Nk_B T/h$ and the required nucleation rate in the MD simulations ($dn/dt \sim 1/\tau_R$). In the present study, we use $\tau_R = 0.2$ ns in MD simulations unless noted otherwise.

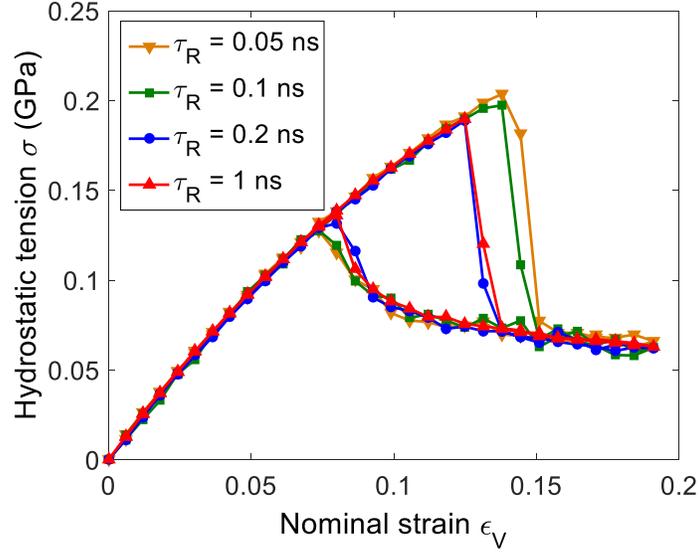

**Figure 3**. Hydrostatic stress-strain diagrams by MD simulations with different relaxation times ($T = 300$ K and $N = 8000$).

## 4.2 Effect of initial liquid volume

Another limitation of the MD simulation is the small box size $L_0$ or $V_0 = L_0^3$, the initial volume of the liquid water, which depends on the number of water molecules in the simulation box. By varying the number of molecules in MD simulations, we obtain the hydrostatic stress-strain diagrams for different initial volumes in Fig. 4a ($T = 300$ K). The nonlinear elastic response before cavitation remains unaffected. Onset of cavitation is slightly delayed for the smallest initial volume ($N = 2197$ and $L_0 = 4.05$ nm), but nearly identical for the other three cases ($N = 8000$, 17576 and 32768). The average stress after cavitation depends on the initial volume: a larger initial volume leads to lower tensile stresses and a lower critical strain for de-cavitation during unloading.

By classical nucleation theory, the initial volume of liquid water affects the cavitation pressure through the nucleation rate only, which has been found to be insignificant for the prediction of cavitation pressure (Eq. 10) [1, 2, 18]. In contrast, the modified nucleation theory in Section 3 predicts a more significant effect, not only on the nucleation rate but also on the free energy through Eq. (16). As written in Eq. (11), formation of a cavity increases the surface energy but decreases the elastic strain energy in the liquid. The competition leads to a critical strain for cavitation that depends on the initial volume: a larger initial volume leads to a lower critical strain (Fig. 4c) and correspondingly a lower critical stress (Fig. 4d). Moreover, the radius of cavity increases as the initial volume increases, leading to lower stresses after cavitation as dictated by Eq. (17). To compare with the MD simulations, we show in Fig. 4b the hydrostatic stress-strain diagrams predicted by the modified nucleation theory for the same initial volumes. The agreement is reasonable except for the case with the smallest initial volume, where onset of cavitation in the MD simulation occurs at a strain lower than the predicted critical strain. For the other three cases, while the predicted critical strain decreases with increasing initial volume, onset of cavitation occurs at similar strain levels, higher than the predicted value. Interestingly, the energy barrier $\Delta\Phi_{max}$ in Eq. (19) is found to be similar at the same nominal strain ($\varepsilon_V \sim 0.12$) where the onset of cavitation is observed in the MD simulations. This suggests that cavitation occurs at similar nucleation rates for the three cases. On the other hand, the de-cavitation strain or stress upon unloading is lower but follows the same trend as the predicted critical values.

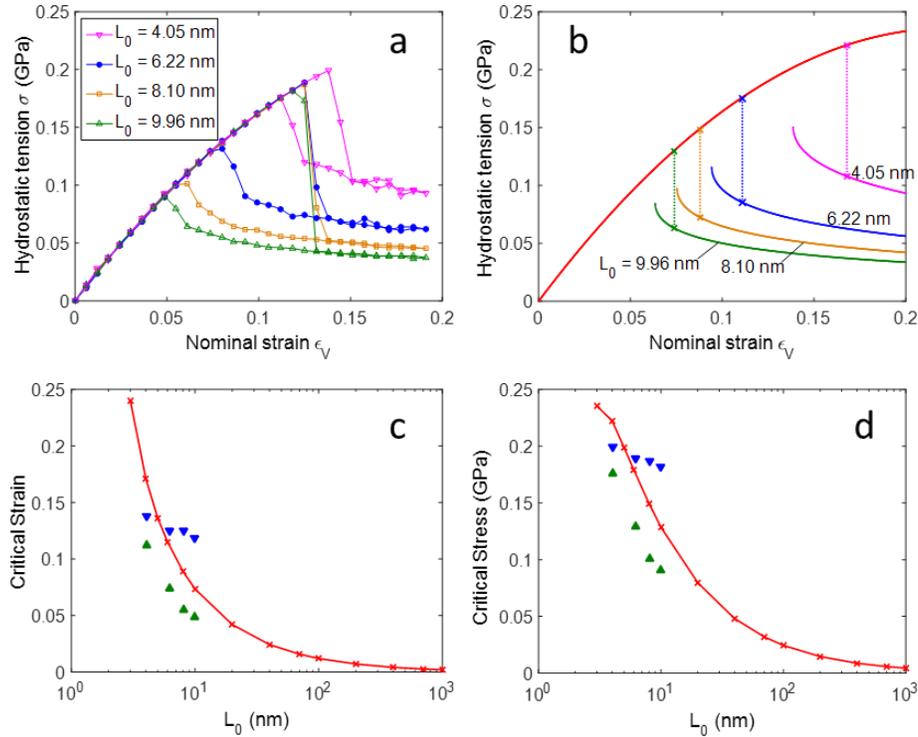

**Figure 4**. (a) Hydrostatic stress-strain diagrams by MD simulations with different simulation box sizes, and (b) diagrams predicted by the modified nucleation theory, with the vertical lines for onset of cavitation. (c) Predicted critical strain and (d) critical stress for cavitation under volume-controlled stretching, in comparison with the upper and lower critical values from MD simulations (filled symbols).

The modified nucleation theory predicts that the critical strain and stress for cavitation decreases with increasing initial volume ($V_0 = L_0^3$). As shown in Figs. 4c and 4d, the MD simulations predict an upper critical strain (stress) for cavitation and a lower critical strain (stress) for de-cavitation. Except for the case with the smallest initial volume, the theoretical prediction lies in between the two critical values from MD, which is expected by qualitative consideration of the nucleation rate. Within the limited volume range by MD simulations, the upper critical strain (stress) appears to be less dependent on the initial volume, while the lower critical strain (stress) follows the same trend as the critical strain predicted by the modified nucleation theory. Extrapolating to much larger initial volumes beyond the MD simulations, the predicted critical strain and stress decrease continuously toward zero. In other words, if the initial volume of liquid water is infinitely large ($V_0 \to \infty$), the critical strain approaches zero and the liquid phase is unstable with respect to cavitation under any hydrostatic tension (or negative pressure). If the vapor pressure is taken into account in the calculation, the liquid phase with $V_0 \to \infty$ would be unstable under any pressure below the equilibrium vapor pressure. This limiting scenario is dictated by classical thermodynamics. However, when the liquid water is confined within a finite volume, it could remain stable without cavitation down to a negative pressure. The strong dependence of the critical strain and stress for cavitation on the initial volume as predicted by the modified nucleation theory may offer a possible explanation for the large discrepancies in the values of the critical negative pressure obtained from experiments [1-11]: a critical stress above 100 MPa could be obtained with a small initial volume ($L_0 < 10$ nm), while the critical stress could be much lower with a larger initial volume, around 30 MPa for $L_0 \sim 100$ nm. It should be noted that strictly volume-controlled stretching has not been realized in experiments. The microscopic Berthelot tube method [4-6] may be the closest to volume-controlled stretching, but it also involves changing temperature, which may have additional effects as discussed next.

### 4.3 Effect of temperature

Cavitation of water at different temperatures is studied by both MD simulations and the modified nucleation theory. Hydrostatic stress-strain diagrams from MD simulations are presented in Fig. 5a for a range of temperatures with the water in its stable liquid phase at atmospheric pressure. The same number of water molecules ($N = 8000$) is used in all simulations. After the initial NPT relaxation at the atmospheric pressure, the equilibrium simulation box size ($L_0$) varies slightly with the temperature (i.e., thermal expansion), as listed in Table I. The corresponding mass density of the liquid water decreases with increasing temperature within the range. Upon stretching, the hydrostatic stress-strain diagram exhibits qualitatively similar behavior at all temperatures considered, namely, a nonlinear elastic rising of the hydrostatic tension followed by an abrupt drop due to cavitation and a slowly decreasing tension after cavitation. Fitting the nonlinear elastic behavior with Eq. (12), we obtain the two elastic moduli that depend on the temperature, as listed in Table I. The linear bulk modulus $K_1$ reaches a maximum at $T = 300\,\text{K}$ and then decreases as the temperature increases, whereas the second-order modulus $K_2$ decreases monotonically with temperature. As a result, the liquid water is less stiff (or more compressible) at higher temperatures ($T > 300$ K). Correspondingly, the cavitation stress in the MD simulations decreases with increasing temperature, although the critical strain is similar for the three cases with $T > 300\,\text{K}$. After cavitation, the average stress decreases slightly with increasing temperature.

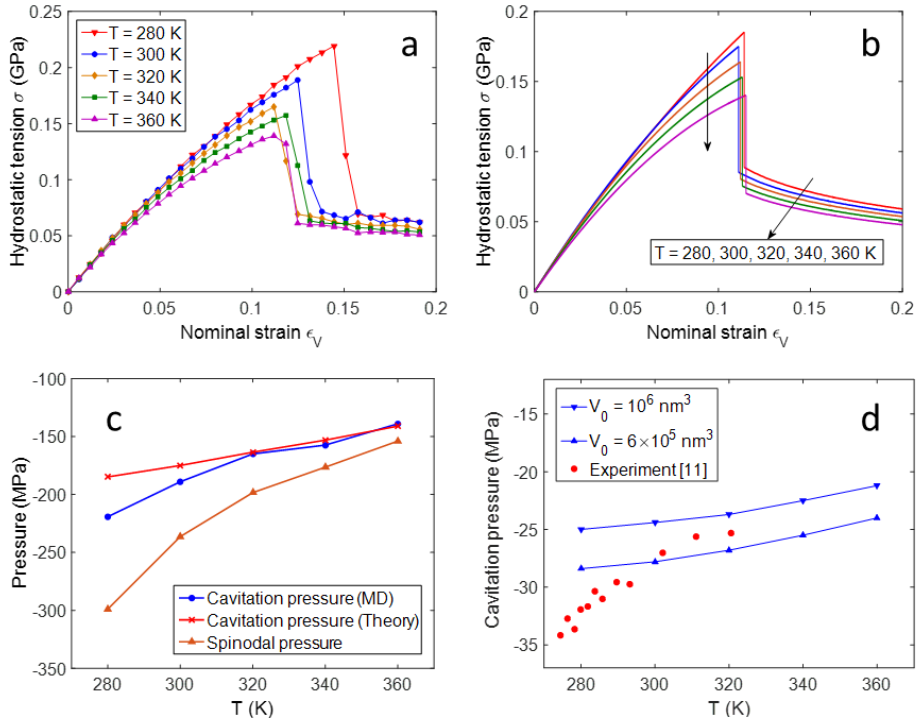

**Figure 5**. (a) Hydrostatic stress-strain diagrams of water at different temperatures by MD simulations ($N = 8000$); (b) Hydrostatic stress-strain diagrams predicted by the modified nucleation theory; (c) Critical pressure for cavitation as a function of temperature from MD and the modified nucleation theory, along with the predicted spinodal pressure. (d) Critical pressure

for cavitation predicted by the modified nucleation theory with relatively large initial volumes, in comparison with the data from acoustic-based measurements [8, 11].

Table I: Temperature dependent parameters obtained from MD simulations using TIP4P/2005.

| $T$ (K) | $L_0$ (nm) | $K_1$ (GPa) | $K_2$ (GPa) | $\gamma$ (N/m) |
|---|---|---|---|---|
| 280 | 6.219 | 2.01 | -3.38 | 0.0722 |
| 300 | 6.223 | 2.09 | -4.62 | 0.0693 |
| 320 | 6.242 | 2.07 | -5.40 | 0.0665 |
| 340 | 6.265 | 1.99 | -5.61 | 0.0632 |
| 360 | 6.293 | 1.88 | -5.74 | 0.0598 |

The prediction by the modified nucleation theory depends on temperature through three parameters: the two elastic moduli, $K_1$ and $K_2$, and the surface tension $\gamma$. We obtain the elastic moduli from the MD simulations and use the following equation to obtain the surface tension at various temperatures:

$$\gamma(T) = c_1 \left(1 - \frac{T}{T_c}\right)^{11/9} \left[1 - c_2 \left(1 - \frac{T}{T_c}\right)\right]. \tag{20}$$

This equation is used by the International Association for Properties of Water and Steam to describe the experimental values of the surface tension of water [30]. The parameters in Eq. (20) were determined from MD simulations using the TIP4P/2005 potential for water [30], and they are: $c_1 = 0.22786\,\text{N/m}$, $c_2 = 0.6413$, and $T_c = 641.4\,\text{K}$, which are used here to obtain the surface tension in Table I. With the temperature-dependent parameters, the modified nucleation theory predicts the hydrostatic stress-strain diagrams at different temperatures in Fig. 5b. At each temperature, a transition from the homogeneous liquid phase to the cavitated state is predicted at a critical strain under volume-controlled stretching, and the hydrostatic stress drops abruptly at the transition. Compared to the MD simulations in Fig. 5a, the modified nucleation theory accurately predicts the critical stress for the three cases with $T > 300\,\text{K}$, but under-predicts the critical stress at lower temperatures, as shown in Fig. 5c. This may be qualitatively understood based on the nucleation rate in Eq. (19). Since the nucleation rate generally increases with increasing temperature, onset of cavitation could occur slightly above the critical strain at relatively high temperatures but would require a larger strain to lower the energy barrier ($\Delta\Phi_{max}$) at relatively low temperatures. Remarkably, the critical strain for cavitation predicted by the modified nucleation theory is similar (~0.11) at all temperatures considered here, but the critical stress depends on temperature (Fig. 5c) due to the temperature-dependent elastic moduli. After cavitation, the average stress decreases with increasing temperature largely due to the temperature-dependent surface tension, while the radius of cavity varies slightly with temperature.

The critical negative pressure ($p_c = -\sigma_c$) for cavitation is shown in Fig. 5c as a function of temperature, from both MD simulations and the modified nucleation theory. In addition, the nonlinear hydrostatic stress-strain relation in Eq. (12) may be extrapolated to predict a spinodal pressure ($p_s = -\sigma_s$) by setting $d\sigma/d\varepsilon = 0$, which leads to: $p_s(T) = K_1^2/(4K_2)$, as shown in Fig. 5c. The spinodal pressure has been predicted theoretically as a stability limit for liquid water under negative pressure [12, 38]. Interestingly, two kinds of temperature dependence have been obtained previously for the spinodal pressure, one with a minimum at around 320 K [12] and the other increasing monotonically with temperature [38]. Apparently, with the elastic moduli obtained from the MD simulations using TIP4P/2005 in the present study, the predicted spinondal pressure increases monotonically with temperature, consistent with other MD simulations [3, 38, 39]. While theoretically significant, the spinodal pressure in principle cannot be measured directly in experiments because of cavitation ($p_s < p_c$). In MD simulations, cavitation may be suppressed by using a small number of water molecules and as a result, the hydrostatic stress-strain diagram can be obtained up to the spinodal limit [38, 39]. It has been suggested that measurements of the cavitation pressure would help to settle the fundamental debate on the spinodal pressure and the associated anomalies in the phase diagram of water [1, 11]. However, measurements of cavitation pressure in pure water have also been challenging. One set of experiments using the microscopic Berthelot tube method [4-6] have reached a critical negative pressure of -140 MPa at 315 K. Other experimental methods have consistently yielded values of around -30 MPa at room temperature [7-11]. The discrepancy may be understood by the modified nucleation theory. Unlike classical nucleation theory, the modified nucleation theory as proposed in this study predicts a strong dependence of the cavitation pressure on the volume of water (Fig. 4d). With a small initial volume ($V_0 = 700$ nm$^3$), the cavitation pressure of -140 MPa is predicted at 300 K. On the other hand, with a large initial volume ($V_0 = 6\times10^5$ nm$^3$), the cavitation pressure at 300 K is predicted to be -28 MPa. Using relatively large initial volumes, we obtain the cavitation pressure as a function of temperature from the modified nucleation theory as shown in Fig. 5d, for comparison with the data from acoustic-based experiments [8, 11]. This illustrates the potential to resolve the discrepancies in the cavitation pressure of water measured by various experiments.

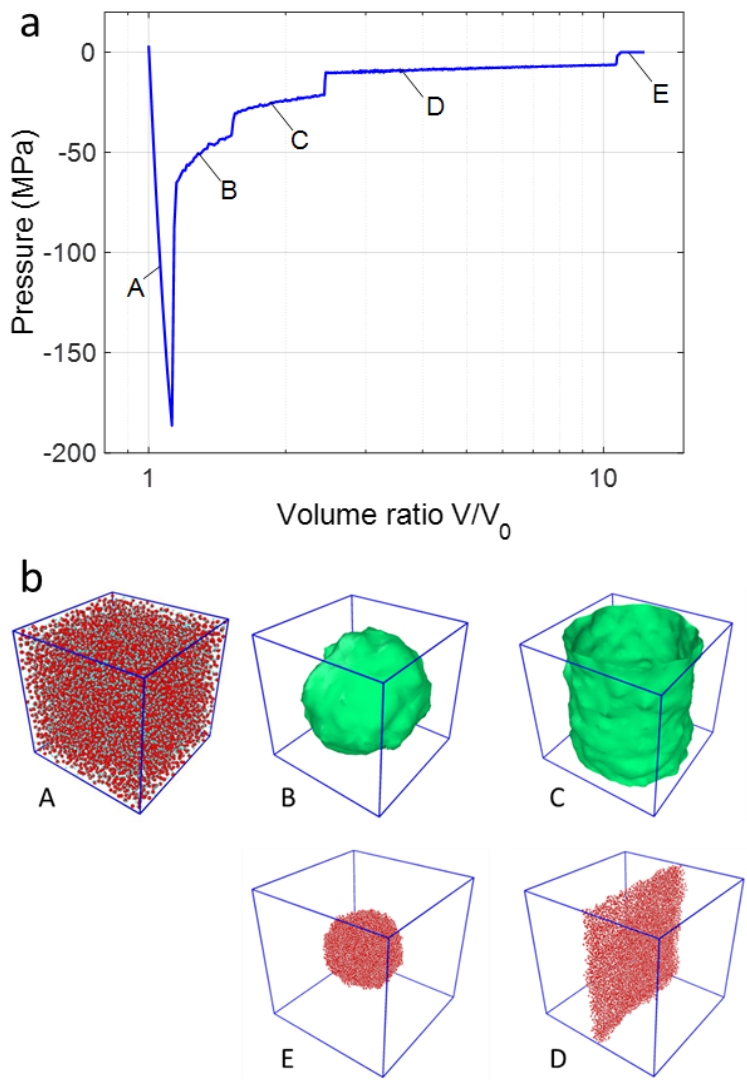

**Figure 6**. (a) Pressure-volume diagram of water under volume-controlled stretching by MD simulation ($T = 300$ K and $N = 8000$). (b) A sequence of morphological transitions from spherical cavity (B) to cylindrical (C) cavity and alternating layers (D), and to a liquid droplet (E). Water molecules in the liquid phase are shown in A, D and E, while the cavity surface is shown in B and C.

## 4.4 An isothermal process of liquid-vapor phase transition

The most common liquid-vapor phase transition occurs under the isobaric condition, changing the temperature at a constant pressure, while the specific volume changes abruptly before and after the transition. The two phases may co-exist at the saturation temperature, forming a saturated liquid-vapor mixture. Alternatively, the liquid-vapor phase transition may occur under the isothermal condition, by changing the pressure at a constant temperature. In this case, the two phases co-exist at the saturation pressure that depends on temperature. Yet another scenario of liquid-vapor phase transition may be achieved by isothermal volume-controlled stretching as considered in this study. With the total volume constrained, the liquid water (of a small initial volume) could remain stable with respect to cavitation under a negative pressure, far below the saturation pressure (~3 kPa for water at 300 K). Upon cavitation, the pressure rises abruptly, but still below the saturation pressure. After cavitation, the liquid and vapor phases co-exist with an interface, forming a heterogeneous mixture, where the pressure difference in the two phases is balanced by surface tension. It may be expected that the volume-averaged pressure would continue increasing by further stretching and eventually become positive in the region of a stable, homogeneous vapor phase, thus completing the liquid-vapor phase transition through an isothermal stretching process. Indeed, such a process can be simulated by MD as illustrated in Fig. 6. With $N = 8000$ for the number of water molecules and $T = 300$ K, we stretch the simulation box continuously to more than 10 times of the initial volume. Interestingly, the hydrostatic stress-strain diagram or equivalently, the pressure-volume diagram (Fig. 6a) exhibits three subsequent transitions after cavitation, each with a notable jump of the pressure. Further examination reveals that each transition is associated with a morphological change of the water in the simulation box as shown in Fig. 6b. First, the spherical cavity grows as the total volume increases and becomes unstable when its diameter is almost the side length of the simulation box, leading to a transition of the cavity shape from spherical to cylindrical. Such a geometric transition is necessary to accommodate the increasing volume fraction of the cavity (vapor phase) within the cubic simulation box. Next, with further stretching, the cylindrical cavity grows and becomes unstable again when its diameter reaches the side length of the simulation box, leading to another transition as the vapor cylinders in neighboring boxes (with periodic boundary conditions) coalesce to form a layered structure with alternating liquid and vapor phases. Such a layered structure, with a liquid layer sandwiched between two vapor layers, is similar to the slab model used by Vega and Miguel [30] to calculate the surface tension of water. Subsequently, the liquid layer is being stretched equi-biaxially and its thickness decreases. The liquid layer becomes unstable when it is too thin and breaks up to form a spherical droplet. At this stage, the volume fraction of liquid is small and the average pressure is nearly zero. If stretched further (not performed in the present study due to computational limitations), the droplet would shrink in size and eventually vanish, leaving behind a homogeneous vapor phase in the simulation box. Therefore, the isothermal liquid-vapor phase transition can be completed by volume-controlled stretching, through a sequence of morphological transitions with co-existing liquid/vapor phases including spherical and cylindrical cavities as well as liquid droplets.

## 5. Summary

We conduct MD simulations to study cavitation of water under volume-controlled stretching. Motivated by the MD simulations, we propose a modified nucleation theory for cavitation, taking into account the nonlinear elastic compressibility of liquid water. The numerical and theoretical results are compared and discussed to elucidate the effects of volume and temperature on cavitation. A few key findings are summarized as follows.

- Liquid water exhibits a nonlinear elastic behavior under hydrostatic tension and remains stable within the confined volume until onset of cavitation.
- By the modified nucleation theory, cavitation is predicted as a first-order phase transition with a critical strain and a corresponding critical stress (or negative pressure).
- Unlike classical nucleation theory, the modified nucleation theory predicts a strong dependence of the critical strain and stress for cavitation on the initial liquid volume, which may offer a possible explanation for the discrepancies in the values of the critical negative pressure obtained from experiments.
- Both MD simulations and the modified nucleation theory predict that the critical stress for cavitation decreases with increasing temperature, which may be attributed to the temperature-dependent elastic moduli (compressibility) and surface tension.
- By MD simulations, we illustrate an isothermal process of liquid-vapor phase transition under volume-controlled stretching, with a sequence of morphological transitions from spherical to cylindrical cavities and to a liquid droplet.

**Acknowledgements**

PW and RH gratefully acknowledge financial support of this work by the National Science Foundation through Grant No. CMMI-1562820.